# Magnetic lattices for ultracold atoms and degenerate quantum gases

Yibo Wang · Prince Surendran · Smitha Jose ·
Tien Tran · Ivan Herrera · Shannon Whitlock ·
Russell McLean · Andrei Sidorov · Peter Hannaford

**Abstract** We review recent developments in the use of magnetic lattices as a complementary tool to optical lattices for trapping periodic arrays of ultracold atoms and degenerate quantum gases. Recent advances include the realisation of Bose-Einstein condensation in multiple sites of a magnetic lattice of one-dimensional microtraps, the trapping of ultracold atoms in square and triangular magnetic lattices, and the fabrication of magnetic lattice structures with sub-micron period suitable for quantum tunnelling experiments. Finally, we describe a proposal to utilise long-range interacting Rydberg atoms in a large spacing magnetic lattice to create interactions between atoms on neighbouring sites.

**Keywords** Magnetic lattices, ultracold atoms, degenerate quantum gases, quantum simulation

Yibo Wang, Prince Surendran, Smitha Jose, Tien Tran, Russell McLean, Andrei Sidorov, Peter Hannaford
Centre for Quantum and Optical Science, Swinburne University of Technology, Melbourne, Victoria 3122, Australia

Ivan Herrera
Dodd-Walls Centre for Photonic and Quantum Technologies
Department of Physics, University of Auckland,
Private Bag 92019, Auckland, New Zealand

Shannon Whitlock
Physikalisches Institut, Universität Heidelberg, Im Neuenheimer Feld 226, 69120 Heidelberg, Germany

P. Hannaford
email: phannaford@swin.edu.au

## 1 Introduction

Since the advent of laser cooling and trapping techniques in the 1980s and 90s [1-3], optical lattices produced by interfering laser beams have become an indispensable tool for trapping periodic arrays of ultracold atoms and degenerate quantum gases (see, e.g., [4-6] for reviews). Applications include quantum simulations of condensed matter phenomena [6], trapping of atom arrays in high precision atomic clocks [7] and quantum gas microscopes [8], and the realisation of quantum gates for quantum information processing [9, 10]. These 'artificial crystals' allow precise control over system parameters, such as the lattice geometry, inter-particle interaction and lattice perfection, and, in principle, provide an ideal platform to achieve almost perfect realisations of a variety of condensed matter phenomena (e.g., [5, 6]). Examples include the superfluid to Mott insulator transition [11], the metal to insulator cross-over in honeycomb lattices [12], the Ising spin model for a 1D spin chain [13], the Hubbard model involving antiferromagnetic correlations [14], low-dimensional quantum systems [15, 16], disordered systems involving Anderson localisation [17, 18], topological edge states and the quantum Hall effect [19], and arrays of Josephson junctions [20].

An alternative approach for producing periodic lattices for trapping ultracold atoms involves the use of arrays of magnetic microtraps created by patterned magnetic films [21-36], current-carrying conductors [37-40], nano-magnetic domain walls [41], vortex arrays in superconducting films [42] or pulsed gradient magnetic fields [43, 44]. In the present mini-review we focus on recent developments in the trapping of ultracold atoms in magnetic lattices of microtraps based on patterned magnetic films. These magnetic lattices have a high degree of design flexibility and may, in principle, be tailored with nearly arbitrary configurations and lattice spacing [29] without restrictions imposed by optical wavelengths. In addition, magnetic lattices do not require (intense) laser beams, they are free of spontaneous emission, they have relatively little technical noise or heating, and they involve state-selective atom trapping allowing radio-frequency (RF) evaporative cooling to be performed in the lattice and RF spectroscopy to be used to characterise the trapped atoms *in situ* [45, 46]. Finally, magnetic lattices are well suited for mounting on atom chips and incorporating into devices such as 'atomtronic' circuits [47]. However, magnetic lattices are still in their infancy compared with optical lattices, due largely to the difficulty of fabricating suitable magnetic microstructures with well controlled potentials, especially lattices with sub-micron periods suitable for quantum tunnelling.

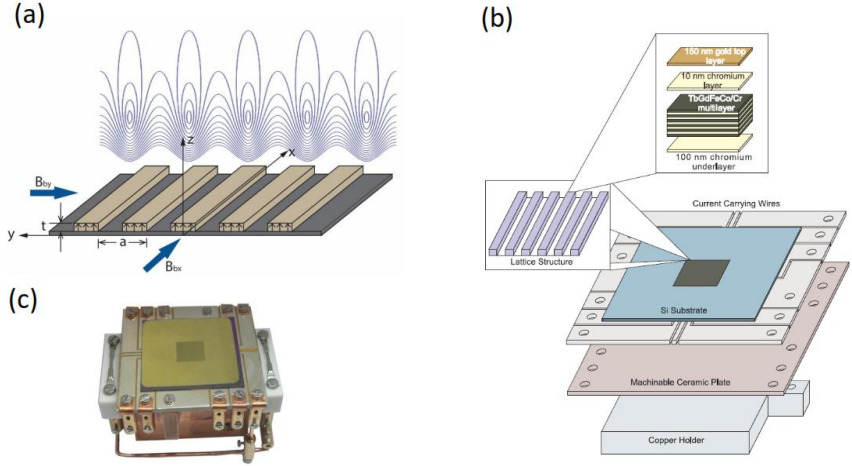

**Fig. 1** (**a**) Magnetic lattice of 1D microtraps produced by an array of perpendicularly magnetised magnets with period *a* and bias fields $B_{bx}$, $B_{by}$ [58]. The contour lines are magnetic equipotentials calculated for typical parameters. (**b**) Construction of the magnetic lattice atom chip. The machined silver foil beneath the magnetic lattice structure contains U-shaped and Z-shaped current-carrying conductors for trapping and loading the atoms. Reproduced from Singh et al. [26] with the permission of IOP Publishing. (**c**) Photograph of the magnetic lattice atom chip coated with a reflecting gold film [59]

In this article we review recent progress in the development of magnetic lattices based on patterned magnetic films for trapping periodic arrays of ultracold atoms and degenerate quantum gases, and discuss future prospects for the application of magnetic lattices.

## 2 One-dimensional magnetic lattices

Magnetic lattices consisting of arrays of one-dimensional microtraps are a useful testing ground prior to progressing to more complex two-dimensional geometries. They are a natural extension of the magnetic mirrors [48] proposed by Opat et al. in 1992 [49] and subsequently realised using arrays of permanent magnets [50-52] and current-carrying conductors [53-55]. A magnetic mirror may be turned into a magnetic lattice of 1D microtraps by applying a uniform bias field to interfere with the rotating magnetic field of the periodic array (Fig. 1(a)), as described by Eq (1) below.

For an infinite periodic array of long magnets in the *x-y* plane with perpendicular magnetisation $M_z$, periodicity $a$ and bias fields $B_{bx}$, $B_{by}$, the magnetic field components at distances $z \gg a/2\pi$ from the bottom of the magnets are given approximately by [22]

$$[B_x; B_y; B_z] \approx [B_{bx}; B_0 \sin(ky)e^{-kz} + B_{by}; B_0 \cos(ky)e^{-kz}] \quad (1)$$

where $k = 2\pi/a$, $B_0 = 4M_z(e^{kt} - 1)$ is a characteristic surface magnetic field (in Gaussian units), and $t$ is the thickness of the magnets. The magnetic field minimum (or trap bottom) $B_{min}$, trapping height $z_{min}$, barrier heights $\Delta B_{y,z}$, and trap frequencies $\omega_{y,z}$ for an atom of mass $m$ in a harmonic trapping potential are given by

$$B_{min} = |B_{bx}| \quad (2)$$

$$z_{min} = \frac{a}{2\pi} \ln\left(\frac{B_0}{|B_{by}|}\right) \quad (3)$$

$$\Delta B_y = \left(B_{bx}^2 + 4B_{by}^2\right)^{\frac{1}{2}} - |B_{bx}| \quad (4a)$$

$$\Delta B_z = \left(B_{bx}^2 + B_{by}^2\right)^{\frac{1}{2}} - |B_{bx}| \quad (4b)$$

$$\omega_y = \omega_z = \omega_{rad} = \frac{2\pi}{a}\left(\frac{m_F g_F \mu_B}{m|B_{bx}|}\right)^{1/2}|B_{by}| \quad (5)$$

where $m_F$ is the magnetic quantum number, $g_F$ is the Landé g-factor and $\mu_B$ is the Bohr magneton. $B_{min}$, $z_{min}$, $\Delta B_{y,z}$ and $\omega_{y,z}$ may all be controlled by adjusting the bias fields $B_{bx}$ and $B_{by}$. Equations (1) – (5) illustrate how the characteristics of the magnetic lattice can be varied by varying the bias fields $B_{bx}$, $B_{by}$ and how the magnetic lattice can be switched on or off by switching $B_{by}$ on or off. Equations (2) – (5) are useful for providing scalings for the various parameters.

In 2005 Sinclair et al. [56] created a periodic array of 1D magnetic traps, or magnetic 'waveguides', made from a sinusoidal magnetisation pattern of period 106 μm written on videotape plus bias fields, and successfully produced a single Bose-Einstein condensate (BEC) in one of the waveguides. In 2007 Boyd et al. [57] created an array of 1D traps produced by a hard disk platter written with a periodic pattern of period 100 μm plus bias fields, and produced a condensate in one of the traps.

In 2008 Singh et al. [26] produced a 10 μm-period magnetic lattice of 1000 1D traps formed from a perpendicularly magnetised 1 μm-thick TbGdFeCo film on a grooved silicon substrate on an atom chip plus bias fields (Fig. 1(b)). About $10^8$ $^{87}$Rb atoms were initially trapped in a mirror magneto-optical trap (MOT) and then confined in a compressed MOT using the quadrupole field from a current-carrying U-wire plus bias field. Atoms in the $|F = 2, m_F = +2\rangle$ low field-seeking state were then transferred to a Z-wire magnetic trap where they were RF evaporatively cooled to ~15 μK [60].

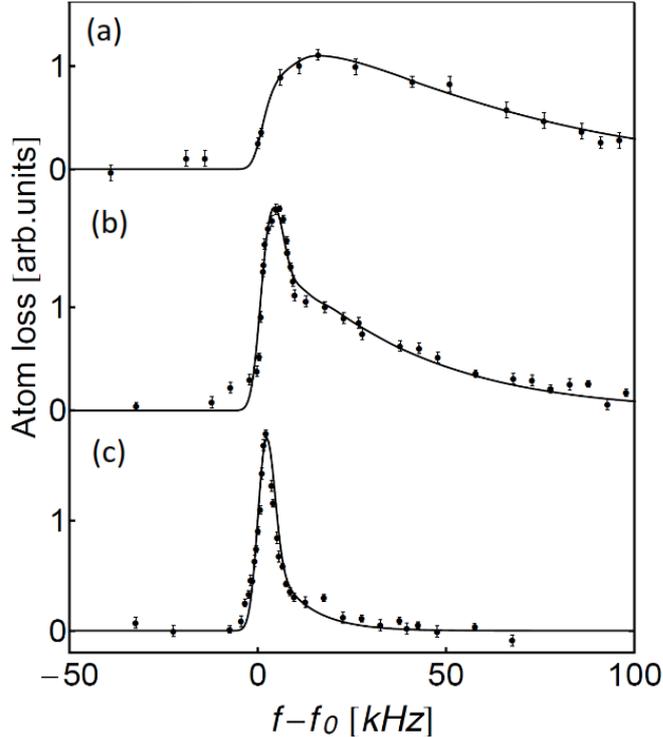

**Fig. 2** Radiofrequency spectra of $^{87}$Rb $|F=1, m_F=-1\rangle$ atoms in one of the ~100 atom clouds trapped in a 10 μm-period 1D magnetic lattice, demonstrating the onset of Bose-Einstein condensation with increased evaporative cooling. The solid lines are fits to the data points based on a self-consistent mean-field model for a BEC plus thermal cloud as described in the text. The trap depths $\delta f = f_f - f_0$, temperatures and atom numbers obtained from this analysis are (**a**) 600 kHz, 2.0 μK, 5350 atoms, (**b**) 400 kHz, 1.3 μK, 3430 atoms and (**c**) 100 kHz, 0.38 μK, 200 atoms. Adapted from [35]

The atoms in the Z-wire trap were then brought close (~5 μm) to the chip surface by ramping down the Z-wire current ($I_z$) and ramping up the bias field $B_{by}$ to 30 G to create the magnetic lattice microtraps. When the Z-wire trap merged with the magnetic lattice traps, $I_z$ was reduced to zero with $B_{bx} = B_{min} = 15$ G. In this way, typically $3 \times 10^6$ atoms were loaded into ~100 magnetic lattice traps in the central region of the lattice, with barrier heights ~1 mK and trap frequencies in the range $\omega_{rad}/2\pi = 20 - 90$ kHz, $\omega_{ax}/2\pi \approx 1$ Hz. Radiofrequency spectroscopy measurements indicated temperatures $> 150$ μK, which were limited largely by the weak axial confinement that prevented efficient evaporative cooling in the lattice.

### 3 Bose-Einstein condensation in multiple magnetic lattice sites

A significant breakthrough was made in recent experiments by Jose et al. [33] and Surendran et al. [35] using the above magnetic lattice chip. The $^{87}$Rb atoms were optically pumped into the $|F=1, m_F=-1\rangle$ low field-seeking state to minimise losses due to three-body recombination in the tightly confining magnetic traps [61, 62] and additional axial confinement was applied, with trap frequencies in the range $\omega_{rad}/2\pi = 1.5 - 20$ kHz, $\omega_{ax}/2\pi = 260$ Hz. Figure 2 shows radiofrequency spectra recorded for one of the ~100 atom clouds trapped in the magnetic lattice as the atoms were evaporatively cooled to lower trap depths $\delta f = f_f - f_0$ and lower temperatures (where $f_f$ is the final evaporation frequency and $f_0$ is the trap bottom). The spectra show the evolution from a broad thermal cloud distribution (Fig. 2(a)) to a bimodal distribution characteristic of a thermal cloud plus a narrow BEC distribution (Fig. 2(b)) to an almost pure BEC distribution (Fig. 2(c)) as the atom clouds are cooled through the critical temperature (1.6 μK for an ideal gas with $N = 3000$ atoms).

The fits to the data points in Fig. 2 are based on a self-consistent mean-field model for a BEC plus thermal cloud convolved with a Gaussian magnetic noise function (FWHM=4.3 kHz) [35]. The model includes the repulsive interaction among atoms in the BEC and in the thermal cloud and the mutual interaction between them but neglects the kinetic energy of the condensate atoms via the Thomas-Fermi approximation and the effects of gravity sag in the tight magnetic traps. The fits provide measurements of the trap bottom ($f_0$), atom temperature ($T$), condensate fraction ($N_C/N$) and chemical potential ($\mu$). The effect of temperature is to change both the width of the broad thermal cloud component and the fraction of atoms in the condensate.

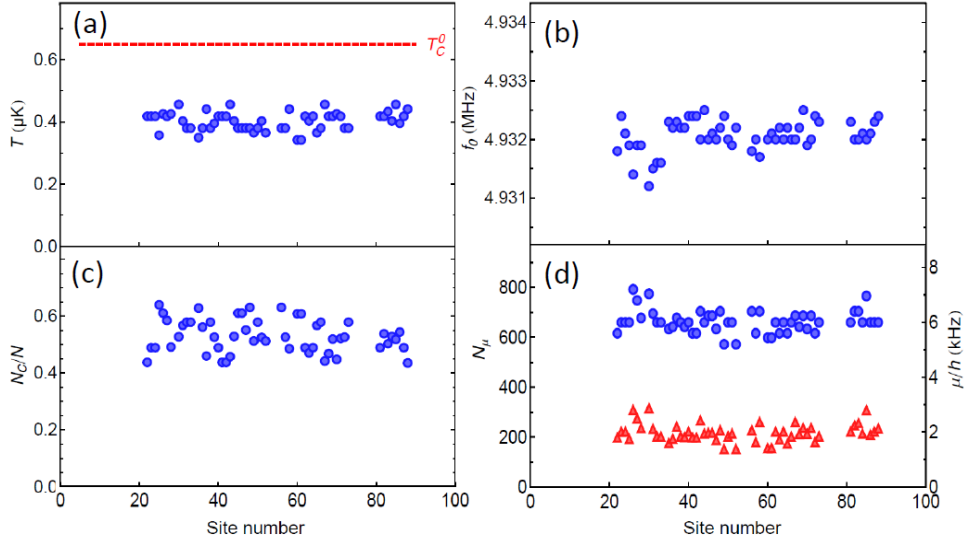

**Fig. 3** (**a**) Atom temperature $T$, (**b**) trap bottom $f_0$, (**c**) condensate fraction $N_C/N$ and (**d**) chemical potential $\mu$ (blue circles) and atom number $N_\mu$ derived from $\mu$ (red triangles), determined from fits to the RF spectra for multiple sites across the central region of the magnetic lattice, with trap depth $\delta f = 100$ kHz, $\omega_{rad}/2\pi = 7.5$ kHz. The red dashed line in (**a**) represents the ideal-gas critical temperature for 220 atoms. Adapted from [35]

Radiofrequency spectra taken simultaneously for all atom clouds across the central region of the magnetic lattice showed similar bimodal distributions to Fig. 2 with site-to-site variations in the above quantities that were consistent with the measurement errors (Fig. 3). In particular, the trap bottom $f_0$, which could be determined precisely from measurements of the frequency at which there were no atoms lost (Fig. 2(c)), showed one-sigma variations of only $\pm 0.3$ kHz (or $\pm 0.4$ mG) in 5 MHz (Fig. 3(b)), reflecting the high uniformity in the central region of the magnetic lattice. The atom temperature is well below the ideal-gas critical temperature for all sites (Fig. 3(a)) and large condensate fractions are observed for all sites (Fig. 3(c)).

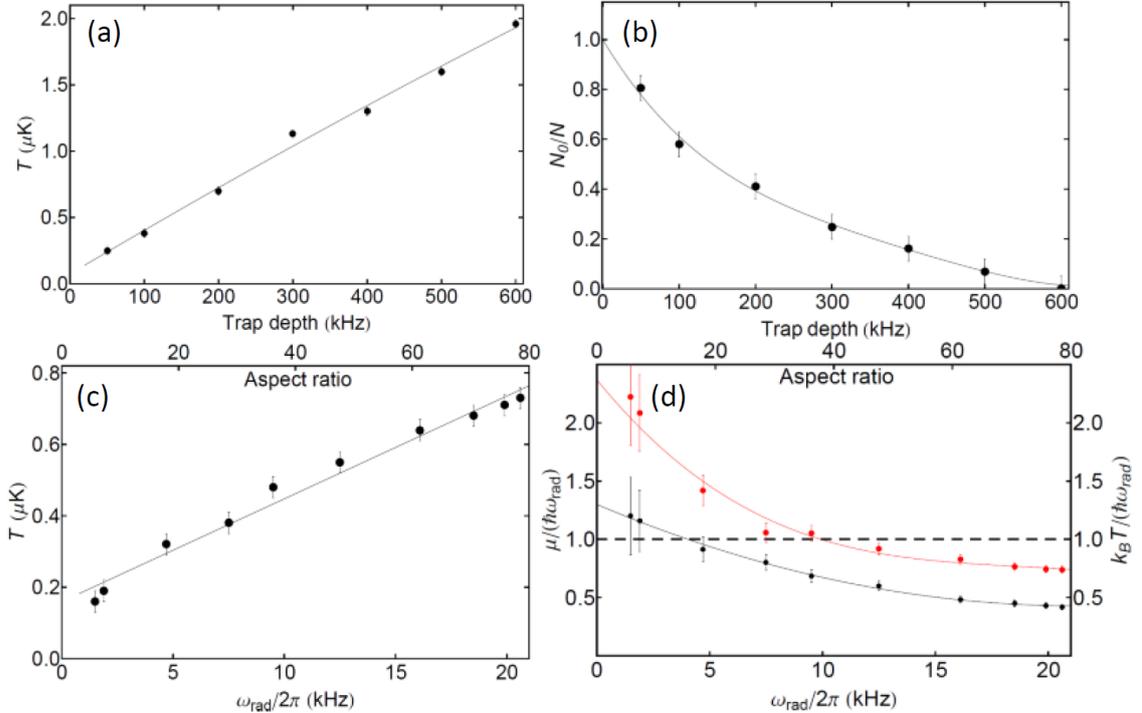

**Fig. 4** Variation of (**a**) atom temperature $T$ and (**b**) condensate fraction $N_C/N$ with trap depth $\delta f$ at trap frequencies $\omega_{rad}/2\pi = 7.5$ kHz, $\omega_{ax}/2\pi = 260$ Hz; and variation of (**c**) atom temperature and (**d**) $\mu/\hbar\omega_{rad}$ (black points) and $k_BT/\hbar\omega_{rad}$ (red points) with radial trap frequency $\omega_{rad}$ at $\delta f = 100$ kHz. The horizontal dashed line in (**d**) corresponds to $k_BT = \mu = \hbar\omega_{rad}$, which represents the energy of the lowest radial vibrational excited state. Adapted from [35]

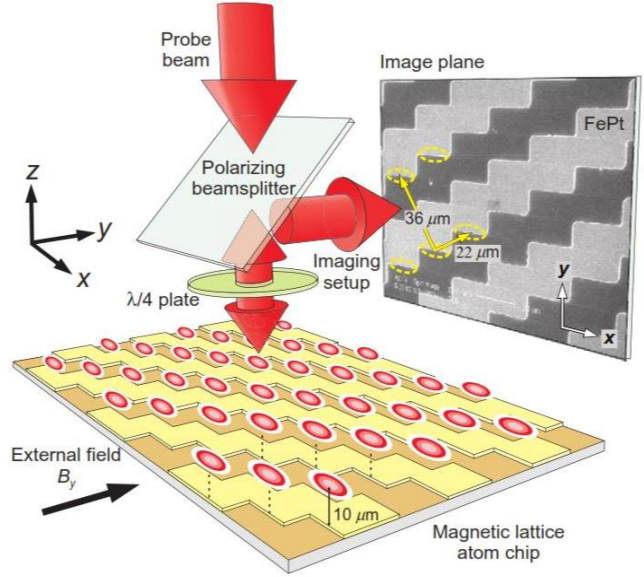

**Fig. 5** Schematic of the absorption imaging of atom clouds trapped in a two-dimensional rectangular magnetic lattice. Reproduced from Whitlock et al. [27] with the permission of IOP Publishing

Figure 4 shows the variation of atom temperature $T$ and condensate fraction $N_C/N$ with trap depth $\delta f$ and radial trap frequency $\omega_{rad}/2\pi$. At the lowest radial trap frequency (1.5 kHz) a temperature of 0.16 μK is achieved in the magnetic lattice (Fig. 4(c)) and at the smallest trap depth (50 kHz) a condensate fraction of ~80% is observed (Fig. 4(b)). For radial trap frequencies $\omega_{rad}/2\pi > 10$ kHz, both the chemical potential μ and the thermal energy $k_B T$ become smaller than the energy of the lowest radial vibrational excited state $\hbar\omega_{rad}$ (Fig. 4(d)), which represents the quasi-one-dimensional Bose gas regime [15, 63-67]. In Fig. 4(c) the atom temperatures determined from our analysis are seen to increase with radial trap frequency which is provisionally attributed to reduced efficiency of evaporative cooling due to suppression of rethermalising collisions in the quasi-1D regime [65].

Our one-dimensional magnetic lattice is able to create a large number of almost equivalent, highly elongated atom clouds. Thus, these lattices seem well suited for studying one-dimensional quantum gases. Using realistic parameters, e.g., $\omega_{rad}/2\pi = 50$ kHz, $\omega_{ax}/2\pi = 100$ Hz, $N = 50$ atoms and $T = 50$ nK, it should be possible to reach the strongly interacting Tonks-Girardeau regime, which requires the Lieb-Liniger interaction parameter $\gamma \approx 2a_s/(n_{1D} l_{rad}^2) > 1$ and the temperature parameter $t = 2\hbar^2 k_B T/(mg^2) < 1$ [64]. Here, $n_{1D} = N/l_{ax}$, $l_{rad} = \sqrt{\hbar/(m\omega_{rad})}$, $g \approx 2\hbar\omega_{rad} a_s$ and $a_s$ is the s-wave scattering length.

## 4 Two-dimensional magnetic lattices

For many of the applications for which magnetic lattices are attractive, such as the quantum simulation of condensed matter phenomena and quantum information processing, two-dimensional lattices are usually required.

In 2007 Gerritsma et al. [25] created a two-dimensional rectangular magnetic lattice using a patterned, perpendicularly magnetised FePt magnetic film with periods of 36 μm and 22 μm in the $x$, $y$ directions plus bias fields. Whitlock et al. [27] extended this work to load and image individual clouds of $^{87}$Rb $|F = 2, m_F = +2\rangle$ atoms in over 500 sites of the magnetic lattice. Figure 5 shows a schematic of the absorption imaging of the atom clouds trapped in the two-dimensional rectangular magnetic lattice. Losses due to rapid three-body recombination of the $^{87}$Rb $|F = 2, m_F = +2\rangle$ atoms during evaporative cooling in the tight traps prevented the formation of Bose-Einstein condensates with an observable condensate fraction.

One of the challenges with designing a two-dimensional magnetic lattice, especially lattices with high symmetry such as a square lattice [32], is the occurrence of magnetic field zeros, which lead to Majorana spin flips and loss of atoms. In 2010 Schmied et al. [29] developed a numerical algorithm for designing optimised magnetic microstructures to create periodic arrays of microtraps of various geometries with non-zero magnetic field minima. Figures 6(a) and (b) show magnetic film patterns designed to create square and triangular magnetic lattices at a trapping height $z_{min} = a/2$. The corresponding magnetic potentials are shown in Figs. 6(c) and (d). The magnetic film patterns are equivalent to those produced by an electric current passing around the perimeter of the film pattern, which for the square magnetic lattice has a similar shape to that of a square array of Z-wires (which have non-zero magnetic field minima). Schemes for loading the ultracold atoms from a Z-wire trap into a square or triangular magnetic lattice are given in [29, 68]

In 2014 Leung et al. [34] used the above algorithm to design square and triangular magnetic lattices with period 10 μm, and successfully loaded $^{87}$Rb $|F = 2, m_F = +2\rangle$ atoms into both lattices at temperatures of about 35 μK (Fig. 7).

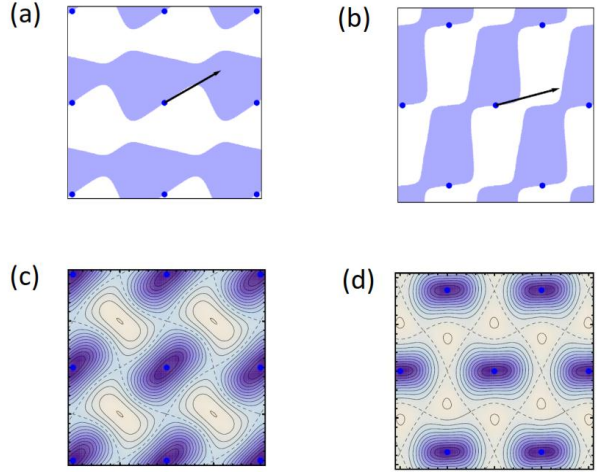

**Fig. 6** Magnetic film patterns to create (**a**) square and (**b**) triangular magnetic lattices at a trapping height $z_{min} = a/2$ [29]. Blue regions represent the magnetic film. The black dots indicate the positions of the magnetic field minima and the black arrows show the direction of the magnetic field at the minima. The resulting magnetic potentials are shown in (**c**) and (**d**), where the blue regions represent the potential minima. Reproduced from Schmied et al. [29] with the permission of IOP Publishing.

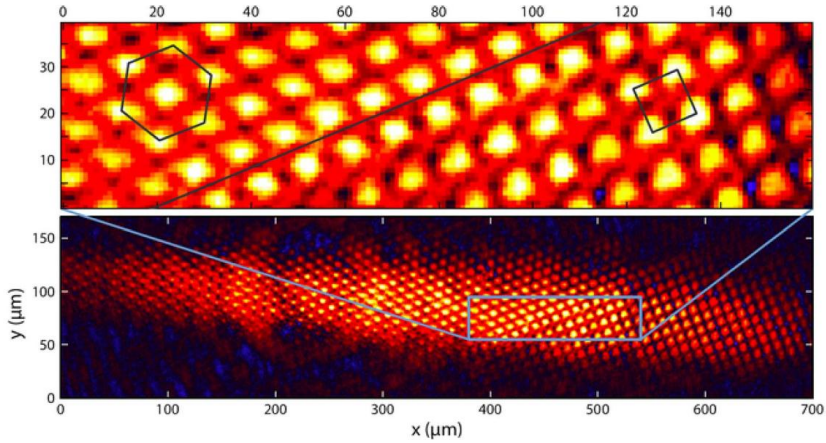

**Fig. 7** Absorption images of $^{87}$Rb atoms loaded into triangular (left side) and square (right side) magnetic lattices with period 10 μm. The lattice sites contain ~360 and ~440 atoms for the square and triangular lattices, respectively. The upper image is a magnified view of the central region. Reproduced from Leung et al. [34] with the permission of AIP Publishing.

## 5 Sub-micron period magnetic lattices

To date, one-dimensional magnetic lattices [26, 30, 33, 35] and two-dimensional rectangular [25, 27], square and triangular [34] magnetic lattices have been produced to trap periodic arrays of ultracold atoms with periods down to 10 μm. For these lattice periods and for realistic barrier heights, there can be no quantum tunnelling of atoms between lattice sites, and the arrays of atoms represent isolated clouds with no interaction between them. To achieve significant tunnelling, lattice spacings of less than 1 μm are required.

Herrera et al. [36] have recently reported the fabrication and characterisation of square and triangular magnetic lattice structures with a period of 0.7 μm. For $a = 0.7$ μm and barrier height $V_0 \sim 12E_R$ (where $E_R = h^2/8ma^2$ is the lattice recoil energy), the tunnelling rate is estimated to be $J \sim 17$ Hz [8]. This rate is compatible with the estimated lifetimes of the trapped atom clouds in these lattices (see below) and is suitable for realising the superfluid to Mott insulator transition, demonstrating the accessibility of magnetic lattices to the Hubbard model.

The magnetic microstructures were fabricated by patterning a Co/Pd multi-atomic layer film (8 bi-layers of 0.28 nm Co + 0.9 nm Pd) [36, 69] on a silicon substrate using electron-beam lithography followed by reactive ion etching. Multi-atomic layer Co/Pd film was chosen because of its strong perpendicular magnetic anisotropy and very small grain size (~ 6 nm [70] compared with ~ 40 nm for TbGdFeCo [71]), allowing smooth and well-defined magnetic potentials at very small periods [72], as well as its high remanent magnetisation (5.9 kG) and coercivity (1.0 kOe) [36].

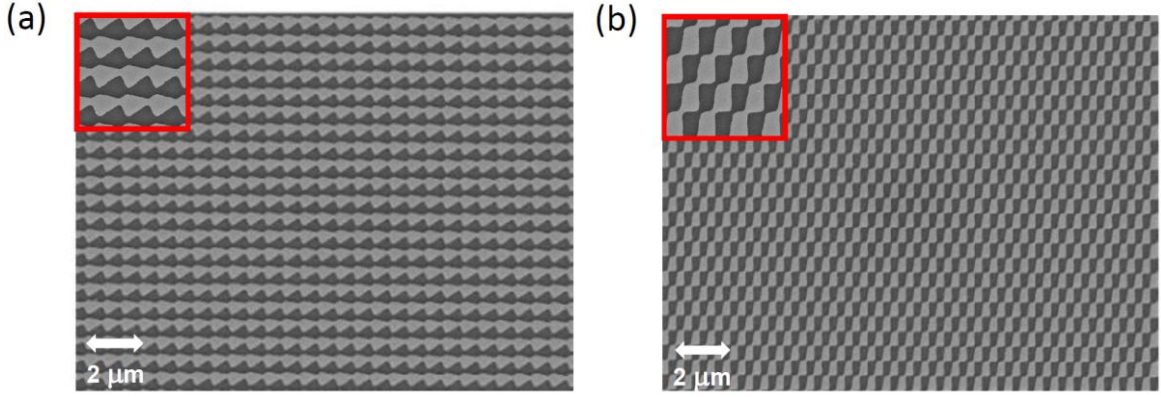

**Fig. 8** SEM images of the fabricated Co/Pd magnetic microstructures to create (**a**) square and (**b**) triangular magnetic lattices with period 0.7 μm. The light grey regions represent the (unetched) magnetic film. The insets show part of the images magnified

Figure 8 shows SEM images of the 0.7 μm-period square and triangular magnetic microstructures, which illustrate the quality of the structures. The magnetised microstructures were characterised using magnetic force microscopy and magneto-optical Kerr effect [36].

Our simulations indicate that the square magnetic lattice with magnetisation $4\pi M_z = 5.9$ kG, magnetic film thickness $t = 2.2$ nm and bias fields $B_{bx} = 1.7$ G, $B_{by} = −0.8$ G create cylindrically symmetric traps at $z_{min}$= 0.35 μm from the magnetic film surface, with the long axis in the [110] direction (Fig. 6(c)). For these parameters the trap minima are $B_{min} = 1.1$ G, the barrier heights are $\Delta B_{y,x}$ = 1.4 G (or 47 μK for $F = 1, m_F = −1$), $\Delta B_z = 0.8$ G (or 26 μK), and the trap frequencies perpendicular and parallel to the long axis are $\omega_\perp/2\pi = \omega_z/2\pi = 120$ kHz, $\omega_\parallel/2\pi = 37$ kHz. Comparable results are obtained for the triangular lattice [36]. Larger barrier heights may be achieved by using thicker Co/Pd magnetic films.

Previous studies have shown that the trap lifetime of an atom cloud may be significantly shortened as the cloud approaches within a few microns of a surface [73-75]. At these distances the attractive van der Waals and Casimir-Polder forces between the atoms and the surface can distort the magnetic potential so that it is no longer trapping. For atoms located 0.3 μm from the surface of a 0.7 μm-period magnetic microstructure, we estimate a critical trapping frequency at which the magnetic traps begin to fold to be $\omega_{crit}/2\pi \approx 44$ kHz [36], which is smaller than the trapping frequency $\omega_z/2\pi = 120$ kHz. In the case of a micro-structured magnetic surface, the attractive van der Waals and Casimir-Polder forces may be swamped by the exponentially increasing repulsive magnetic force close to the surface. The trap lifetime may also be shortened by Johnson magnetic noise which arises from random thermal currents flowing in a conducting surface film that can induce spin flips and loss of atoms [73, 76-78]. For $^{87}$Rb $|F = 1, m_F = −1\rangle$ atoms trapped 0.3 μm from a 50 nm-thick gold conducting film at $T = 300$ K, we estimate a thermal spin-flip lifetime of $\tau_{Au} \approx 180$ ms [36], which is longer than the estimated tunnelling time of 60 ms for a 0.7 μm-period lattice with barrier height $V_0 \sim 12 E_R$ (20 mG), or 13 ms for $V_0 \sim 6\, E_R$. The spin-flip lifetime may be lengthened by using a reflecting film with higher resistivity such as palladium for which $\tau_{Pd} \approx 870$ ms.

## 6 Long-range interacting Rydberg atoms in a magnetic lattice

An alternative approach for creating interaction between atoms on neighbouring sites of a magnetic lattice is to utilise long-range interacting Rydberg atoms [31, 34, 68, 79]. Highly excited Rydberg atoms can be orders of magnitude larger than ground-state atoms, making them very sensitive to fields and to one another. At large separations the interaction between Rydberg atoms is largely due to van der Waals interactions, scaling as $C_6/r^6$, where $C_6$ scales with principal quantum number as $n^{11}$. Each magnetic lattice site is prepared with one Rydberg atom in an ensemble of ground-state atoms via 'Rydberg blockade', in which the presence of the Rydberg atom shifts the energy levels of nearby atoms, suppressing subsequent excitation of other atoms in the ensemble [80]. The characteristic range of the Rydberg-Rydberg interaction is given by the blockade radius, $r_b \approx |C_6/\Omega|$, which for a typical atom-light coupling constant $\Omega/2\pi \sim 1$ MHz is $5 − 10$ μm (depending on the Rydberg state). To prepare a single Rydberg atom in an ensemble of spatial extent $l$ on each site of a lattice of period $a$ requires $l \ll r_b \leq a$, which can be met for a large spacing ($a \sim$ 10 μm) magnetic lattice.

A potential issue when using long-range interacting Rydberg atoms in magnetic lattices is the effect of stray electric fields [79, 81-83]. During each cooling and trapping sequence Rb atoms can stick to the surface of the atom chip to create inhomogeneous electric fields [84] that can perturb the nearby Rydberg atoms. Studies by Tauschinsky et al. [79] of Rb Rydberg atoms trapped at distances down to 20 μm from a gold-coated chip surface have revealed small distance-dependent energy shifts of $\sim \pm 10$ MHz for $n \approx 30$. Recent studies have demonstrated that the stray electric fields can be effectively screened out by depositing a uniform film of Rb over the entire gold surface [81] or by using a smooth monocrystalline quartz surface film coated with a monolayer of Rb adsorbates [82]. More studies are required to understand these effects.

## 7    Summary and perspectives

Significant advances have recently been made in the development of magnetic lattices based on patterned magnetic films as a complementary tool to optical lattices for trapping periodic arrays of ultracold atoms and degenerate quantum gases.

Trapping of ultracold atoms in a one-dimensional magnetic lattice [26, 30] and two-dimensional rectangular [25, 27], square and triangular magnetic lattices [34] with periods down to 10 μm has been demonstrated. Using a recently developed numerical algorithm [29] magnetic lattices based on patterned magnetic films may now, in principle, be tailored with nearly arbitrary configurations and lattice spacings. In the future, it should be possible to produce complex 2D geometries, such as honeycomb and kagome lattices and superlattices. Magnetic lattices based on patterned magnetic films on an atom chip are compact, robust and permanent, making them suitable for incorporating into devices, such as 'atomtronic' circuits [47].

Bose-Einstein condensation has been achieved in multiple sites of a magnetic lattice of one-dimensional microtraps with period 10 μm [33, 35]. High condensate fractions (~80%), low atom temperatures (~0.16 μK) and a high degree of lattice uniformity have been demonstrated in a magnetic lattice.

For the magnetic lattices produced to date, which have periods ≥ 10 μm, the arrays of ultracold atoms represent isolated clouds with no interaction between the atoms on neighbouring sites. To enable quantum tunnelling of atoms between sites, lattices with sub-micron periods are required. High quality square and triangular magnetic lattice structures with periods of 0.7 μm have recently been fabricated by patterning a Co/Pd multi-atomic layer magnetic film [36]. These magnetic lattices would allow the quantum simulation of condensed matter systems such as the Hubbard model [5]. In the future, it should be possible to load fermionic atoms, such as $^{40}$K, into magnetic lattices to simulate electrons in condensed matter systems such as graphene-like systems.

Another scheme to create interaction between lattice sites is to utilise long-range interacting Rydberg atoms in a large spacing (~10 μm) magnetic lattice [31]. These magnetic lattices would allow the quantum simulation of spin models such as the Heisenberg spin model including anisotropic or beyond nearest neighbour spin-spin interactions between Rydberg atoms [85].

**Acknowledgments** This work is supported by an Australian Research Council Discovery Project grant (DP130101160). We thank M. Singh for his contributions to the early stages of our experiments; M. Albrecht and D. Nissen from the University of Augsburg for providing the Co/Pd magnetic films; and A. Balcytis, P. Michaux and S. Juodkazis for fabricating the magnetic microstructures. We thank the Institute of Physics Publishing for permission to reproduce Figs. 1(b), 5 and 6 and the American Institute of Physics Publishing for permission to reproduce Fig. 7.

**Conflict of interest** The authors declare that they have no conflict of interest.